\documentclass[aps,prper,showpacs,twocolumn,floats,superscriptaddress,10pt]{revtex4-1}
\usepackage{graphicx}
\usepackage{dcolumn}
\usepackage{bm}
\usepackage{amssymb}
\usepackage{nicefrac}

\usepackage[latin3]{inputenc}
\usepackage[makeroom]{cancel}
\usepackage{amsmath}
\usepackage{amsfonts}
\usepackage{amssymb}
\usepackage{color}
\usepackage[left=2cm,right=2cm,top=2cm,bottom=2cm]{geometry}

\usepackage{graphicx} 
\usepackage{dcolumn}
\usepackage{bm}
\usepackage{simplewick}
\usepackage{array}
\usepackage{appendix}

\RequirePackage[
   hyperindex,colorlinks,bookmarksnumbered,
   plainpages=true,pdfstartview=FitH]{hyperref}
\hypersetup{linkcolor=blue,urlcolor=blue,citecolor=blue}
\usepackage{hyperref}

\definecolor{purple}{rgb}{0.5,0,0.6}

\usepackage{ulem}

\begin{document}
\title{Charge Kondo circuit as a detector for electron-electron interactions\\ 
in a Luttinger Liquid}
\date{\today}

\author{T. K. T. Nguyen}
\email{nkthanh@iop.vast.vn}
\affiliation{Institute of Physics, Vietnam Academy of Science and Technology, 10
Dao Tan, 118000 Hanoi, Vietnam}
\author{A. V. Parafilo}
\affiliation{Center for Theoretical Physics of Complex Systems, Institute for Basic
Science, Expo-ro, 55, Yuseong-gu, Daejeon 34126, Republic of Korea}
\author{H. Q. Nguyen}
\affiliation{Institute of Physics, Vietnam Academy of Science and Technology, 10
Dao Tan, 118000 Hanoi, Vietnam}
\author{M. N. Kiselev}
\affiliation{The Abdus Salam International Centre for Theoretical Physics, Strada
Costiera 11, I-34151, Trieste, Italy}

\begin{abstract}
We investigate the effects of the electron-electron interactions on the quantum transport through a charge Kondo circuit. The setup consists of a quantum dot sandwiched between two leads by two nearly transparent single mode quantum point contacts. The size of the interacting area $L$ in the Luttinger liquid
formed in the vicinities of the narrow constrictions is assumed to be
much smaller compared to the size of the quantum dot $a$.
We predict that the interplay between the electron-electron interactions in the Luttinger liquid and the fingerprints of the non-Fermi liquid behavior in the vicinity of the two channel Kondo intermediate coupling fixed point allows one to determine the interaction strength through the power-law temperature scaling of the electric conductance.
\end{abstract}

\maketitle

The phenomenon associated with a resonance scattering of itinerant electrons on a quantum impurity with non-zero
magnetic moment is known as the Kondo effect \cite{Kondo}. While this effect has a long history in connection with the condensed matter community \cite{Hewson}, it keeps attracting a growing interest for several directions of an inter-disciplinary research \cite{QCDKondo1,QCDKondo2,Kondo_cold_atom,Kondo_topo}.
Having many facets related to the effects of strong electron-electron interactions, the Kondo effect appears to be an important player in the physics of both heavy fermion compounds and high temperature superconductors \cite{Hewson}. With the rapid development of nano-technologies the Kondo effect has been
engineered in different nano-devices, where artificial structures can be exploited in the regimes that are inaccessible in the bulk systems with magnetic impurities \cite{Kouwenhoven,GG_1998,Kouwenhoven_98,Goldhaber-Gordon_Nature_2007,Kondo_nanotube1,Kondo_nanotube2,Kondo_molecules1,Kondo_molecules2,Revival_kondo}.

It is known that quantum effects play an important role in thermodynamic and transport properties of nano-structures \cite{Blanter,Kisbook}. A single-electron transistor (SET) which for example can be built of a small (few hundred nanometers size) semiconductor quantum dot (QD) sandwiched between two metallic electrodes \cite{Devoret}, is one of the most elementary nanodevices to investigate quantum effects on transport properties. The electrons' transport in a SET is governed by the Coulomb blockade (CB) phenomenon \cite{Ingold_Nazarov}, which exhibits the electrostatic repulsion between electrons in the small confined region \cite{CB1,CB2,CB3}. When the number of electrons in the QD is odd, at sufficiently low temperature, the transport in SET is featured by the strong correlations, and the Kondo effect is observed \cite{Kouwenhoven,Kouwenhoven_98,GG_1998,Goldhaber-Gordon_Nature_2007}. Namely, below a characteristic temperature which is called the Kondo temperature $T_K$, the unpaired electron in the dot hybridizes with the cloud of conduction electrons in the leads. This produces  a sharp peak of the width $T_K$ (Abrikosov-Suhl resonance  \cite{Hewson}) in the local density of states of the SET. The Kondo effect in a SET in which the QD plays effectively the role of magnetic impurity, is similar to the conventional Kondo effect in the bulk metals \cite{comment1}. 

The high controllability of the nanostructures and fine-tunability by the external electric and magnetic fields applied to the QDs \cite{Blanter,Kisbook} allow one to investigate  several regimes of the Kondo physics in which the devices show different universal behaviors. The SETs in experiments \cite{Kouwenhoven,Kouwenhoven_98,GG_1998} operate in the regime of the conventional single impurity $S=1/2$ single channel Kondo (1CK) model. The systems are characterized by the Kondo temperature $T_{K}$ which determines the universal Fermi liquid (FL) behavior at $T<T_{K}$.  The properties of the Kondo system in the limit $T>T_{K}$ can be accessed through the perturbation theory approach \cite{Kondo,Hewson}.  For an $M$-orbital spin-$S$ Kondo model the way the mobile electrons screen the impurity spin determines the system properties at low temperatures $T<T_{K}$ (here $T_K$ is the Kondo temperature defined for the $M$-orbital spin-$S$ system using a spirit of the 1CK). For instance, for the full screened case in which $M=2S$ the system belongs to the FL universality class \cite{Nozieres_Blandin_1980,CZ1998,Landau}. The physics is almost the same when the system is in underscreened ($M<2S$) regime while the unscreened part of the impurity magnetic moment produces the Curie-type contribution to the physical observables. However, for the overscreened case ($M>2S$) the system is characterized by the non-Fermi liquid (NFL) properties \cite{Nozieres_Blandin_1980,CZ1998,TW1983,AFL1983,AD1984,TW1984,
TW1985,FGN_1,FGN_2,Affleck}.  The NFL regime of the two-channel $S=1/2$ Kondo model has been observed in the experiment \cite{Goldhaber-Gordon_Nature_2007}.

As the Kondo devices are the two component systems consisting of the itinerant and localized subsystems, the effects of interactions are characterized by different energy scales. The Coulomb interaction in the localized subsystem (QD) is quantified in terms of
the charging energy $E_{C}=e^{2}/2C$ where $C$ is the capacitance of the dot. The electron-electron interactions in the itinerant subsystem (mobile electrons) depend on the electron's density and normally are treated as a perturbation in the theory of FL \cite{Landau}. However, the situation is different in one dimension (1D), where the arbitrary weak interactions completely modify the ground state and drive the system from the FL to a Tomonaga-Luttinger liquid (TLL) \cite{Tomonaga,Luttinger} regime. The low-energy excitations in TLL are collective charge and spin density modes. The Hamiltonian describing the interacting 1D electrons is thus mapped onto a noninteracting bosonic Hamiltonian for distinct spin and charge degrees of freedom with effective Fermi velocities and Luttinger parameters \cite{FL_LL1,FL_LL2,gogolin,giamarchi}.
The TLL is characterized by power-law decay of various correlation functions with exponents depending on the strength of the electron-electron interactions through the Luttinger parameter $g$  \cite{Haldane,Solyom}, where $g=1$ for the non-interacting system.
The first and famous example of the effects of electron-electron interactions on the transport in a 1D electron liquid with a single- and double-barrier defect was elucidated by Kane and Fisher \cite{kanefisher}. It was shown that the sign of the interaction determines the perturbative relevance of the impurity to the TLL. Namely, the strength of the barrier(s) grows for repulsive interaction ($g < 1$), while it decreases for the attractive interaction ($g > 1$). Therefore, the conductance vanishes at the impurity site at temperature $T\rightarrow 0$ in the first case and reaches the maximal allowed by quantum mechanics value in the second case. At finite temperature the conductance depends on the temperature in a power-law form.  The Kane-Fisher theory has implications for many quantum systems which are realizable in experiments such as quantum Hall materials \cite{QHmaterial}, electronic quantum circuits \cite{e_circuit, fqh1} and cold atomic gases \cite{cold_atomic, isolated_system1,isolated_system2}.
The influence of electron-electron interactions in the TLL with magnetic impurity on the Kondo correlations has also been addressed in a number of works \cite{Kondo__TLL1,Kondo_TLL2,Kondo_TLL3,Kondo_TLL4,Kondo_TLL5,exp_Kondo_TLL}. It is shown that the exponents in the power-law temperature dependency of the quantum thermodynamic and transport observables are functions of the Luttinger parameters.

The charge Kondo effect \cite{flensberg,matveev,furusakimatveev,pierre2,pierre3}, in contrast to the spin Kondo effect \cite{Hewson,Kouwenhoven,GG_1998,Kouwenhoven_98,Goldhaber-Gordon_Nature_2007}, is associated with the two-fold degeneracy of the charge degree of freedom at the Coulomb peaks {\color{black}and} theoretically described in terms of an iso-spin \cite{thanh2018,thanhprl}. {\color{black}A} charge Kondo setup is implemented by a large metallic Coulomb blockaded  QD strongly coupled to one (or several) lead(s) through an (or several) almost fully transmitting single-mode quantum point contact(s) [QPC(s)]. This setup is described by the theoretical model proposed by Flensberg, Matveev, and Furusaki (FMF) \cite{flensberg,matveev,furusakimatveev}. Recently, two- and three-channel charge Kondo circuits have been investigated
in the breakthrough experiments in semiconductor nanostructures in the integer quantum Hall (IQH) regime \cite{pierre2,pierre3}. These experiments provide successful realization  of the FMF model, \cite{flensberg,matveev,furusakimatveev} with the chiral edge states scattered at QPCs. The two degenerate charge states of the QD in the charge Kondo setup represent the internal degree of freedom of the 
``quantum impurity''. The electron location (namely, in- or out- of QD) is treated as an iso-spin variable \cite{thanh2018,thanhprl} similarly to the iso-spin of the FMF model. The number of QPCs is thus equivalent to the number of the orbital channels in the conventional $S=1/2$ Kondo problem.  Moreover, the new experimental ideas \cite{pierre2,pierre3} enrich the possibilities of getting an access to the multichannel Kondo physics and therefore investigating the NFL regimes \cite{Nozieres_Blandin_1980}.

The electrons transferring through almost transparent QPC in the FMF model are described by a 1D Hamiltonian. In addition,  the charging energy $E_C$ in the experiments \cite{pierre2,pierre3} is much smaller compared to the Fermi energy of the leads. It allows us to linearize the spectra of the 1D electrons near the Fermi points. The Abelian bosonization technique \cite{FL_LL1,FL_LL2,gogolin,giamarchi}, which enables one to treat the {\color{black}electron-electron} interactions in 1D systems exactly, is applied to solve the problems.
The FMF model has been extensively studied for several years \cite{flensberg,matveev,furusakimatveev,thanh2018,
thanhprl,furusakimatveevprb,andreevmatveev,LeHur2002,thanh2010}. However,  mainly due to substantial computational difficulties the 1D electron systems at the QPCs have been assumed noninteracting. The effects of electron-electron interactions in the vicinity of the QPC(s) have been first accounted for very recently \cite{thanh2020,Anton2021}. This motivates us to raise an important question: is it possible to detect the electron-electron interactions in the two-channel charge Kondo (2CCK) circuit built with a Luttinger liquid component(s)?

\begin{figure}
\begin{tabular}{c}
\includegraphics[width=1\columnwidth]{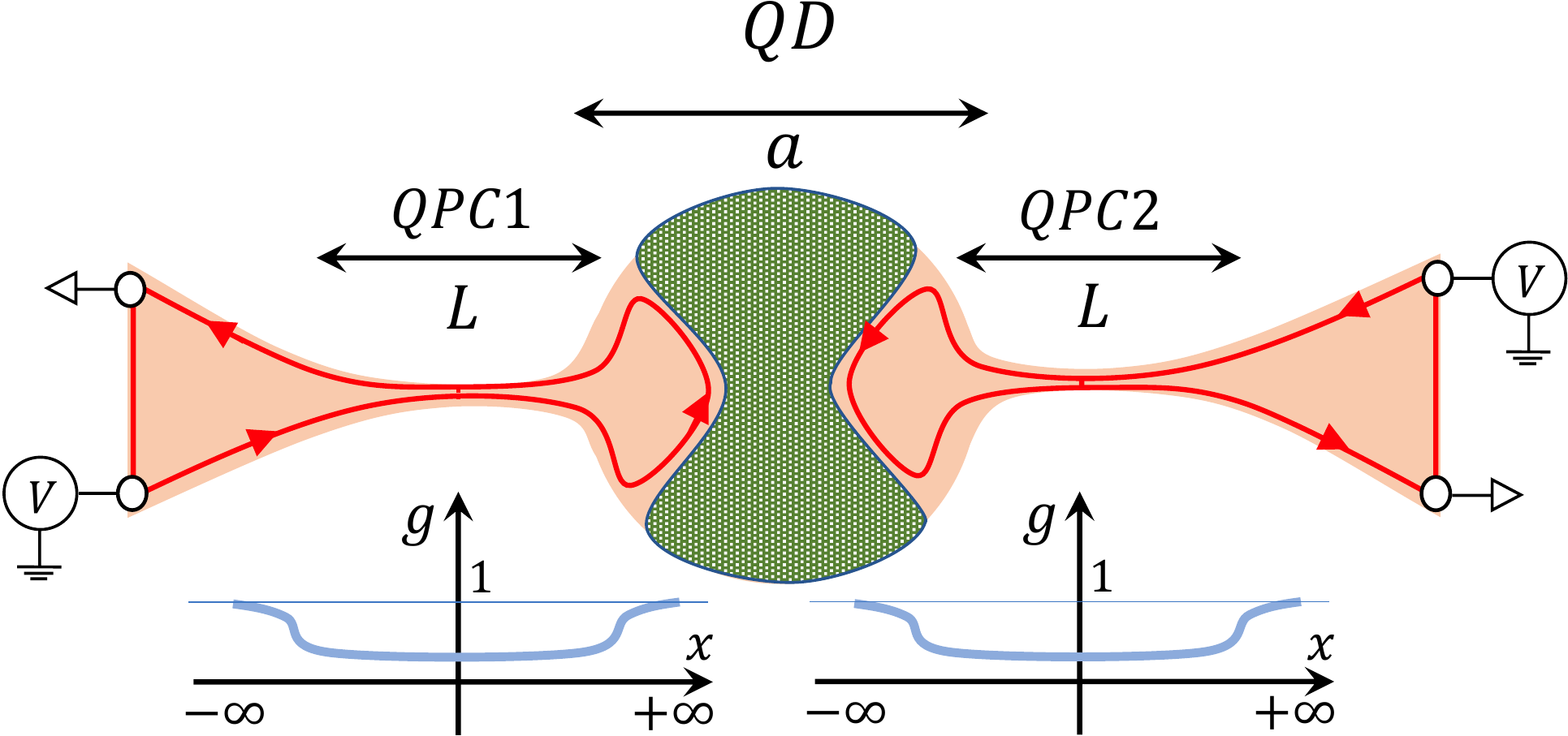}\tabularnewline
\end{tabular}\caption{Schematic setup of two channel charge Kondo circuit
following experiments \cite{pierre2,pierre3}.  The two dimensional electron gas, 2DEG (orange area) is in the integer quantum Hall regime $\nu =2$.
Only one (outer) chiral edge mode (red solid lines with arrows) contributes to the backscattering at the almost transparent single-mode quantum point contacts (QPCs).
A central metallic
island (quantum dot, QD, green hatched area)  is connected to two leads through two uncorrelated QPCs.  As the QPCs cross-talk through the QD, two independent origins
(see insert) are chosen in the middle of each QPC.
The size of QD $a\gg L$ where $L$ is the size of the interacting area. The inhomogeneous 
Luttinger parameters $g(x)$ characterizes the strength of the electron-electron interactions. The interactions asymptotically vanish far away from the QPC 
($x=\pm\infty$).
}
\label{Fig1}
\end{figure}

In this Letter we answer the above question by investigating theoretically the effects of the electron-electron interactions on the linear charge conductance in the multi- QPCs  FMF setup (see Fig.  \ref{Fig1}.). For this purpose the model corresponding to the experimental implementation of Refs. \cite{pierre2,pierre3} is used.  We consider the charge Kondo circuit
simulator shown in Fig. \ref{Fig1}. {\color{black}We assume that a large metallic island (QD) is embedded into two-dimensional electron gas (2DEG) and connects to two large electrodes through the single-mode QPCs.}  The regime of a strong spin polarizing magnetic field is considered and therefore the spin index of the electrons is omitted (see details and discussions  in Refs. \cite{furusakimatveev,furusakimatveevprb}).  The electron-electron interactions in the vicinities of the QPCs are controlled by the external gate voltages \cite{glazman92,exp_interaction_control1,exp_interaction_control2} and play a significant role on the transport properties. We assume that the effects of electron-electron interactions are negligible far away from the narrow constrictions.
The interacting electrons at the 1D edge are described by the TLL model. We investigate the limit $L\ll a$,  where $L$ is the size of the interacting area (TLL) and $a$ is the size of the QD. The perturbative result (in terms of the electron backscattering amplitudes {\color{black}$|r_{1,2}|$} at the {\color{black}QPC$_{1,2}$}) for the linear electric conductance  $G$ (in $\hbar=k_B=1$ units) is given by
\begin{eqnarray}
G & = & G_0\left[1-|r_{+}|^{2}C_{1}\left(g\right)\left(\frac{gE_{C}}{D}\right)^{2g-2}\left(\frac{T}{gE_{C}}\right)^{g-2}\right.\nonumber \\
 &  & \left.-|r_{-}|^{2}C_{2}\left(g\right)\left(\frac{gE_{C}}{D}\right)^{2g-2}\left(\frac{T}{gE_{C}}\right)^{g}\right].\label{eq:general_G_case}
\end{eqnarray}
We denote $|r_{\pm}|^{2}=\left[|r_{1}|^{2}+|r_{2}|^{2}\pm2|r_{1}||r_{2}|\cos\left(2\pi N\right)\right]$, $N$ is a dimensionless parameter (integer at Coulomb valleys and half-integer in CB peaks) controlled by the  gate voltage, $G_0=e^2/4\pi$,  $C_{1}\left(g\right)=(2\gamma)^{g}\pi^{-3/2}\Gamma\left[g/2\right]/4\Gamma\left[1/2+g/2\right]$, and $C_{2}\left(g\right)=g(2\gamma)^{g}\pi^{5/2}\Gamma\left[1+g/2\right]/16\Gamma\left[3/2+g/2\right]$, with $\gamma=e^{C},\;C\approx0.5772$ is the Euler's constant, and $\Gamma\left(x\right)$ is the Gamma function.  The notations $T$, $E_{C}$, and $D$ are used for the temperature, charging energy of the QD, and band-width of the  TLLs, respectively.

The central result of this Letter is given by Eq.~(\ref{eq:general_G_case}). In particular,  a new regime of quantum conductance inaccessible by the Kane-Fisher theory \cite{kanefisher} is predicted.  Equation (\ref{eq:general_G_case}) quantifies the effects of interplay between electron-electron interactions and the Kondo phenomenon in the charge Kondo simulators. The models illustrated by the electric circuits in Fig. \ref{Fig1} can be mapped onto the 2CCK models when two charge states of the QD are degenerate and the mean level spacing is negligibly small. The condition for achieving the intermediate coupling two channel Kondo (2CK) fixed point, at which the physical observables of the system are characterized by the NFL behavior, is the exact symmetry of the two QPC reflection amplitudes: $|r_{1}|=|r_{2}|=|r|$. Applying these conditions to Eq.~(\ref{eq:general_G_case}) we obtain
$G=G_{0}\left[1-4C_{2}(g)|r^{\ast}|^{2}(T/gE_{C})^{g}\right]$
with $|r^{\ast}|=|r|(gE_{C}/D)^{g-1}$. {\color{black} This power-law temperature dependence of the electric conductance accounts for both effects of electron-electron interactions and 2CK correlations.} The conductance approaches its unitary limit $G_{0}$ at
sufficiently low temperatures as $G=G_{0}\left[1-(T/T^{\ast})^{g}\right]$
with $T^{\ast}=gE_{C}/(4C_{2}(g)|r^{\ast}|^{2})^{1/g}$. {\color{black} It allows us to estimate the crossover (Kondo) temperature $T^{\ast}$ and the temperature scaling of the conductance in the TLL-based 2CCK circuits.} The noninteracting case $g=1$ where $G_{0}-G\propto T/T^{\ast}$ at $T\ll T^{\ast}$ has been recently investigated in the breakthrough experiment with the IQH setup \cite{pierre2}. 
Explicit dependence of the electric conductance temperature behavior on the Luttinger parameter $g$ allows one to determine the effects of repulsive electron-electron interactions along the edge through the zero bias anomaly transport measurements \cite{comment_finite_size_QPC}.

Below we summarize the effects of the electron-electron interactions in the QPC-QD-QPC structure on the differential (zero bias) conductance. On the one hand, the electron-electron interactions induce the power-law temperature dependence as $G_{0}-G\propto T^{g-2}$. This effect is similar to the behavior of the linear conductance in the infinite TLL with two weak impurities (potential barriers) \cite{kanefisher,FurNa2}. However, it is notably different from corresponding Kane-Fisher \cite{kanefisher} dependence of the conductance in the TLL with a single impurity $G_{0}-G\propto T^{2g-2}$ \cite{kanefisher,giamarchi}. On the other hand, the $G_{0}-G\propto T^{g-2}$ feature dominates away from the CB peaks ($N$ is half-integer) provided that the third term in Eq.~(\ref{eq:general_G_case}) $\propto C_{2}\left(g\right)T^{g}$ is negligibly small compared to the second term $\propto C_{1}\left(g\right)T^{g-2}$. The validity of Eq.~(\ref{eq:general_G_case}) obtained as the second order perturbation is limited by the validity of the perturbation theory. As it is known from the Kane-Fisher theory \cite{kanefisher},
the fourth order correction ($\propto|r_{1,2}|^{4}$) can play an important role when $g<1/2$ (see Refs. \cite{kanefisher, FurNa2} for details).
Therefore, we assume that the condition $g>1/2$ is satisfied in all equations {\color{black}of this Letter. Moreover, the renormalization of the weak barrier potential at the QPCs $|r_{1,2}^{\ast}|=|r_{1,2}|(gE_{C}/D)^{g-1}$ due to the electron-electron interactions in the TLL reveals the Kane-Fisher phenomenon \cite{kanefisher} and can be analysed with the help of the RG method.  The weak barrier potential's  renormalization shows that if the interaction in the TLL is repulsive
($g<1$), {\color{black}$|r_{1,2}^{\ast}|$} increases when $g$ is decreased reaching the weak coupling limit {\color{black}$|r_{1,2}^{\ast}|\rightarrow 1$}. In contrary, the scattering at the QPC becomes irrelevant ({\color{black}$|r_{1,2}^{\ast}|\rightarrow0$}) if $g>1$ (the attractive interaction regime). In order to satisfy the condition for the perturbation theory calculations performed assuming
the smallness of the renormalized reflection
amplitudes {\color{black}$|r_{1,2}^{\ast}|$}, the electron-electron interactions must be considered being relatively weak. The weak repulsive interaction
in the TLL found in experiments \cite{exp_interaction_control1,experiment2_g}
corresponds to the regime $0.6\leqslant g<1$.}
The non-interacting limit $g=1$ recovers the (A9) of Ref.\cite{furusakimatveevprb}.  We suggest that the domain of parameters accessible for the observation of Eq. (\ref{eq:general_G_case}) can be, for instance:
the Luttinger parameter regime $0.5<g\le 1$, the small charging energy in comparison with the bandwidth of the TLL $E_{C}/D\sim0.2$, and the small reflection amplitudes $|r_{1}|^{2}=|r_{2}|^{2}\sim0.1$. The appropriate temperature range is discussed in the next paragraph.

In the absence of the backscattering at the QPCs, the current operator commutes with the charging Hamiltonian. The conductance is thus insensitive to the Coulomb blockade. When $E_{C}=0$, the equivalent electric circuit for the setup in Fig. \ref{Fig1} is given by a series of two QPCs, where each QPC behaves as a resistor with the resistance $2\pi/e^{2}$. Therefore, the conductance in the limit $E_{C}=0$ is temperature independent. While the electron-electron interactions
result in the renormalization of the conductance quantization in the infinite TLL as $gG_{0}$ \cite{kanefisher,giamarchi}, the unitary limit $G_{0}$ in the finite system described by Eq.~(\ref{eq:general_G_case}) is independent of this interaction. It is consistent with the physics of the finite TLL attached to the FL leads \cite{Maslov,ponomarenkoleads,safileads,exp_finite_wire,FurNa3}. In fact, the interaction immunity of the conductance quantization is the result of two effects, namely, the interplay between the TLL conductance and the contact conductance at the interface of the TLL and the FL reservoir. The contact conductance originates from the processes that take place outside of the interacting area as explained in the Landauer theory on transport in mesoscopic devices \cite{Landauer}.  When $E_{C}\neq0$, the small corrections to the linear conductance
[the second and third terms in Eq.(\ref{eq:general_G_case})] characterize the mesoscopic  Coulomb blockade oscillation (the term $\propto\cos[2\pi N]$).

The fact that the second term of Eq.~(\ref{eq:general_G_case}) dominates when the gate voltage $N$ is detuned from the half-integer values leads to the breakdown of the perturbation theory at sufficiently low temperatures. Therefore, the condition for temperature to validate the perturbation theory reads as  {\color{black}$|r^{\ast}|^{2/(2-g)}gE_{C}\ll T\ll gE_{C}$}.  
We emphasize that the temperature regime {\color{black} $T\leqslant|r^{\ast}|^{2/(2-g)}gE_{C}$} can not be accessed by perturbative calculations and requires the non-perturbative treatment. {\color{black} An alternative possible method is investigating the tunneling instead of the weak backscattering at the QPCs \cite{Fu1998}.}  

Below we present the sketch of the derivation of the result, Eq.(\ref{eq:general_G_case}) for the limit $L\ll a$ \cite{comment2}. The device shown in Fig. \ref{Fig1} consists of a large central metallic island (QD) connected to two large electrodes through two single-mode QPCs. The two TLLs modeling the  two QPCs are ``communicating'' only through the central area of the QD (green hatched area in Fig \ref{Fig1}). Two TLLs have the same length $L$ which is assumed to be much smaller than the size $a$ of the QD: $L\ll a$. This setup is similar to the one in the Furusaki-Matveev (FM) theory \cite{furusakimatveev,furusakimatveevprb}.

To compute the conductance we apply the path integral formulation of 2CCK described in gross details in Appendix A of  Ref. \cite{furusakimatveevprb}.
The action $S$ consists of  three terms $S=S_{0}+S_{C}+S'$, where $S_{0}$, $S_{C}$, and $S'$ describe the TLL 
(formed at the narrow constrictions), weak Coulomb blockade in the QD, and weak backscattering at the QPCs, respectively. Following FM formalism \cite{furusakimatveevprb}, the field $\phi_{1}(x,t)$ describes {\color{black}charge excitations} in the left lead and the left part of QD, while $\phi_{2}(x,t)$ describes {\color{black}charge excitations} in the right lead and the right part of QD. Since the size of the QD is assumed to be large enough, the energy-level spacing is neglected assuming that the condition $v_{F}/a\rightarrow0$ is applied. 

The  Euclidean  action $S_{0}$ describing two separated TLLs with two independent origins chosen in the middle of each QPC is written as: 
\begin{eqnarray}
S_{0} & = & \frac{1}{2\pi}\sum_{\color{black}j=1,2}\int dx\int_{0}^{\beta}dt\frac{1}{g(x)}\nonumber \\
 &  & \times\left[\frac{[\partial_{t}\phi_{\color{black}j}(x,t)]^{2}}{u(x)}+u(x)[\partial_{x}\phi_{\color{red}j}(x,t)]^{2}\right].
 \label{act0}
\end{eqnarray}
Here, both TLLs are assumed to have the same Luttinger parameter $g(x)$
and effective velocity $u(x)$ and  $\beta=1/T$. We assume that  the non-interacting boundary conditions $g(x)=1$, $u(x)=v_{F}$ are satisfied for $|x|>L/2$ (the FL reservoirs and the inside QD).  The Luttinger parameter  $g(x)=g$ and velocity $u(x)=v$ for the $|x|<L/2$ ($v\approx v_{F}/g$) account for the electron-electron interaction's effects inside the finite-size TLLs. The actions $S_{C}$ and $S'$ are given by 
\begin{equation}
S_{C}=\int_{0}^{\beta}\!\!\!dt\frac{E_{C}}{\pi^{2}}\left[\phi_{2}(0,t)-\phi_{1}(0,t)-\pi N\right]^{2},
 \label{actc}
\end{equation}
\begin{equation}
S'=\frac{D}{\pi}\!\!\int_{0}^{\beta}\!\!\!\!\!dt\left\{ |r_{1}|\cos[2\phi_{1}(0,t)]+|r_{2}|\cos[2\phi_{2}(0,t)]\right\},
\label{actpr}
\end{equation}
where the reflection amplitudes of the QPC1(2) are assumed to be small $|r_{1(2)}|\ll1$.

The conductance through the system is computed in the linear response
regime by applying the Kubo formula \cite{Mahan}. In the zero-frequency
limit the linear conductance is related to the average of the current flowing into the QD through the QPC1 and the current flowing out of the QD through the QPC2.  By introducing new (so-called pseudo-spin and charge) variables: $\phi_{s/c}=[\phi_{2}\pm\phi_{1}]/\sqrt{2}$, we define
the current in terms of bosonic
fields as $\hat{I}=e\partial_{t}\phi_{s}\left(0,t\right)/\sqrt{2}\pi$.
Finally,  performing the Fourier transform $\phi_{s}\left(0,t\right)=\sum_{\omega_{n}}\phi_{s}\left(i\omega_{n}\right)e^{-i\omega_{n}t}/2\pi$,
where $\omega_{n}=2\pi nT$ is Matsubara frequency, we re-write
the Kubo formula as follows:
\begin{equation}
G=-i\frac{e^{2}T}{2\pi^{2}}\lim_{\omega\rightarrow0}{\color{black}\omega}\lim_{i\omega_{n}\rightarrow\omega+i0^{+}}\langle\phi_{s}\left(-i\omega_{n}\right)\phi_{s}\left(i\omega_{n}\right)\rangle.\label{eq:definition}
\end{equation}
{\color{black} Since the action $S$ is Gaussian at $x\neq 0$, one can integrate out all degrees of freedom $\phi_{s,c}(x\neq 0, t)$.} The effective action (see details of the derivation in the Supplemental Material \cite{Supplement}) is given by
\begin{eqnarray}\label{eff}
S_{eff} & = & \frac{1}{2\pi\beta}\sum_{n}\sum_{i=s,c}{G_{\omega_{n}}^{-1}(0,0)\phi_{i}(i\omega_{n})\phi_{i}(-i\omega_{n})}\nonumber \\
 &  & ~~~~~~~~ +S_{C}+S'.
\end{eqnarray}
This action is used for the perturbative calculation of conductance. 
The average $\langle\phi_{s}\left(-i\omega_{n}\right)\phi_{s}\left(i\omega_{n}\right)\rangle$ is evaluated assuming smallness of the  reflection amplitudes at both QPCs. Here $G_{\omega_{n}}(x,x')$ is the Green's function
of the spatially inhomogeneous [due to different {\color{black}values} of the Luttinger parameter $g(x)$ and the velocity $u(x)$ inside and outside the interacting area] TLLs attached to the non-interacting leads. Applying the Maslov-Stone approach \cite{Maslov} we obtain (see the Supplemental Material \cite{Supplement}):
{\color{black}
\begin{eqnarray}
G_{\omega_n}(x,x')  =  \frac{g}{2|\omega_n|}\frac{1}{\kappa_{+}^{2}e^{\frac{|\omega_n|L}{v}}-\kappa_{-}^{2}e^{\frac{-|\omega_n|L}{v}}}\nonumber\\\times\left\{\kappa_{-}^{2}e^{\pm\frac{|\omega_n|(x-x')}{v}}e^{-\frac{|\omega_n|L}{v}}+\kappa_{+}^{2}e^{\frac{\mp|\omega_n|(x-x')}{v}}e^{\frac{|\omega_n|L}{v}}+\right.\nonumber\\ +\left.2\kappa_{+}\kappa_{-}\cosh\left[\frac{|\omega_n|(x+x')}{v}\right] \right\} .
\label{eq:GF}
\end{eqnarray}
}
Here $\kappa_{\pm}=(1\pm g)$, and the sign $\pm$ stands for the
case $x>x'$ and $x<x'$, respectively. {\color{black}The  Green's function Eq.~(\ref{eq:GF}) at $x=x'=0$ approaches asymptotically the value $g/2|\omega_n|$ in the ``high''-frequency (temperature) limit ($\omega_n\gg v/L$), while it reaches $1/2|\omega_n|$ when the frequency (or temperature) is lowered to satisfy the condition $\omega_n\ll v/L$. The crossover of these two regimes occurs at the so-called critical temperature $ T^{cr}\sim v_F/g L$.

In the absence of backscattering ($|r_{1,2}|=0$), Eq.~(\ref{eq:definition}) at the dc limit is determined by the low-frequency regime of Eq.~(\ref{eq:GF}). The conductance thus acquires its unitary value $G_0$. However, a correction to the conductance due to the weak backscattering is obtained by using both high- and low-frequency asymptotics of Eq.~(7) which depend on the temperature differently (the $\omega_n=0$ term vanishes, see \cite{Maslov,FurNa3,Supplement}). Therefore, at the high-temperature regime, namely $T\gg T^{cr}$, we obtain Eq.~(\ref{eq:general_G_case}) (see the detail of calculations in \cite{Supplement}), in which the temperature scalings are governed by the  Luttinger parameter $g$ similarly to the case of spatially homogeneous TLL. 
In the limit $T\ll T^{cr}$, the conductance is no longer temperature dependent. The interaction-induced renormalization in the low temperature limit is cut-off by the finite interaction region ($L$) of the TLL \cite{Maslov}. One thus can qualitatively estimate the linear conductance in the low-temperature regime by replacing $T\rightarrow v/L$ in Eq.~(\ref{eq:general_G_case}), see, e.g., Refs.\cite{Maslov,FurNa3}.
} 

{\color{black}
We finally estimate the temperature $T^{cr}$, at which the crossover between high- and low-temperature regimes occur. Choosing the Fermi velocity of the edge modes in the IQH effect as $v_F\sim 10^5 \textrm{m/s}$  \cite{vel1,vel2} and the length of the QPC as $L\approx 0.6 \mu\textrm{m}$ \cite{QPCl}, we obtain $T^{cr}\approx 1 \textrm{K}$. Therefore, in order to observe the effects discussed after Eq.~(\ref{eq:general_G_case}), the device from Ref. \cite{pierre2,pierre3} can be used with some modifications to increase the charging energy of the QD $E_C\sim 1\div 10 \textrm{K}$ and the electron-electron interaction in the vicinity of QPCs. 
}

In conclusion, we have theoretically investigated the effects of the
electron-electron interactions on the electric transport through a charge Kondo circuit. 
Having been inspired by the experiments \cite{pierre2,pierre3} we analysed the experimentally
relevant case of the interacting area in QPC  being small compared to the size of QD 
$L\ll a$ and investigated the temperature regime {\color{black}$|r^{\ast}|^{2/(2-g)}gE_{C}\ll T\ll gE_{C}$}. 
The power-law temperature dependence of the linear conductance is predicted to be modified by the effects of the electron-electron interactions in the TLL through the Luttinger parameter $g$. Notably, when the system approaches the 2CK intermediate coupling fixed point, the universal temperature scaling of the conductance is fully determined by the NFL behavior $G_{0}-G\propto(T/T^{\ast})^{g}$.  We suggest that the key predictions of this Letter are to be used for measuring the effects of  the electron-electron interactions in the two channel charge Kondo -- IQH circuits.

T.K.T.N. and A.V.P. contribute equally to this work. This research in Hanoi is funded by Vietnam National Foundation for Science and Technology Development (NAFOSTED) under grant number 103.01-2020.05. The work of M.N.K is conducted within the framework of the Trieste Institute for Theoretical Quantum Technologies (TQT).


\begin{thebibliography}{10}
\bibitem{comment1}{\color{black} The Kondo effect in bulk  systems leads to
the enhancement of the backscattering. As the result, the resistance increases
with decreasing the temperature.  The Kondo effect in the nanostructures
in turn results in a maximal overlap of the wave functions of the leads. Therefore 
the electric conductance reaches its maximal value (unitary limit) at zero temperature.}
\bibitem{comment2} The limit $a\ll L$ can be investigated the same way as used in the {\it Letter}.  Using perturbative approach we predict that the electric conductance  has the power-law temperature dependence given by Eq.~(\ref{eq:general_G_case}).
\end{thebibliography}
\end{document}